\documentclass[twocolumn,showpacs,prb]{revtex4}
\usepackage{graphicx}

\begin{document}

\title{
Vortex state in a Fulde-Ferrell-Larkin-Ovchinnikov superconductor 
based on the quasiclassical theory
}

\author{Masanori Ichioka} 
\email{oka@mp.okayama-u.ac.jp}
\author{Hiroto Adachi}
\author{Takeshi Mizushima}
\author{Kazushige Machida}
\affiliation{Department of Physics, Okayama University,
         Okayama 700-8530, Japan}

\date{\today}

\begin{abstract}
We investigate the vortex state with 
Fulde-Ferrell-Larkin-Ovchinnikov (FFLO) modulations   
suggested for a high field phase of ${\rm CeCoIn_5}$. 
On the basis of the quasiclassical Eilenberger theory,   
we calculate the three dimensional structure of pair potentials, 
internal magnetic fields, paramagnetic moments, and electronic states, 
for the $s$-wave and the $d$-wave pairings comparatively. 
The $\pi$-phase shift of the pair potential 
at the FFLO nodal plane or at the vortex core 
induces sharp peak states in the local 
density of states, and enhances the local paramagnetic moment.  
We also discuss the NMR spectrum and the neutron scattering as methods 
to detect the FFLO structure. 
\end{abstract}



\maketitle

\section{Introduction}
\label{sec:introduction}

The Fulde-Ferrell-Larkin-Ovchinnikov (FFLO) state~\cite{ff,lo} 
is an exotic superconducting state expected to appear 
at low temperatures and high fields,  
when the paramagnetic depairing is significant 
due to the Zeeman effect under a magnetic field.
In the FFLO state, 
since the Fermi surfaces for up-spin and down-spin electron bands 
are split by the Zeeman effect,   
Cooper pairs of up- and down-spins acquire non-zero momentum 
for the center of mass coordinate of the Cooper pair,  
inducing the spatial modulation of the pair 
potential.~\cite{machida,tachiki,shimahara,klein,buzdin,adachi,ikeda} 
The possible FFLO state is widely discussed in various research fields, 
ranging from superconductors in condensed matter, 
neutral Fermion superfluids in an atomic cloud,
\cite{mizushima,zwierlein,partridge,machida2006} 
to color superconductivity in high energy physics.\cite{casalbuoni}  

Experimentally, the FFLO state is suggested in a high field phase 
of a quasi-two dimensional (Q2D) heavy Fermion superconductor 
${\rm CeCoIn_5}$.~\cite{bianchi,radovan} 
Anomalous behaviors of the sound wave velocity,~\cite{watanabe} 
thermal conductivity,~\cite{capan}  
penetration depth,\cite{martin}
and the NMR spectrum~\cite{kakuyanagi,kumagai} 
in the high field phase are considered as evidence of  
the FFLO structure. 
There, it is supposed that  nodal planes of the pair potential run 
perpendicular to the vortex lines. 

In theoretical studies, 
many calculations for the FFLO state have been done 
by neglecting vortex structure. 
However, we have to include the vortex structure 
in addition to the FFLO modulation, 
because the FFLO state appears at high fields 
in the mixed states.\cite{buzdin,adachi,ikeda} 
Among the FFLO states, there are two possible spatial modulation 
of the pair potential $\Delta$. 
One is the Fulde-Ferrell (FF) state~\cite{ff} with phase modulation 
such as $\Delta \propto {\rm e}^{{\rm i}Qz}$, 
where $Q$ is the modulation vector 
of the FFLO states.  
The other is the Larkin-Ovchinnikov (LO) state~\cite{lo} 
with the amplitude modulation  such as $\Delta \propto \sin Qz$, 
where the the pair potential shows the periodic sign change, 
and $\Delta=0$ at nodal planes. 
We discuss the case of the LO states in this paper, 
since some experimental~\cite{watanabe,capan,kakuyanagi} 
and theoretical~\cite{buzdin,ikeda} works 
support the LO state (at least in low temperature region)
for  the FFLO states in  ${\rm CeCoIn_5}$. 
In the FFLO vortex state, it is instructive to analyze the role of the 
nodal plane in the LO states, which may give a clue to obtain 
clear evidence of the FFLO states among the experimental data.

When we consider vortex structure in the LO state, 
there are two possible choices of the configuration 
for the vortex lines and the FFLO modulation. 
That is, the modulation vector of the FFLO state is parallel~\cite{tachiki} 
or perpendicular~\cite{shimahara,klein} to the applied magnetic field. 
In our study, the former case is investigated by the quasiclassical 
theory.~\cite{tachiki,klein,klein87,ichiokaS,ichiokaD,ichiokaMgB2} 
In a previous quasiclassical study by Tachiki {\it et al.}~\cite{tachiki}  
for this case, they calculate the structure of the FFLO modulation after 
reducing the FFLO states with Abrikosov vortex lattice to a 
problem of the one dimensional (1D) system along the magnetic field direction. 
There, the radial dependence from the vortex core could not be analyzed, 
since the spatial dependence of vortex is approximately changed to the 
averaged one. 
In our study, fully three dimensional (3D) structures of the vortex and 
the FFLO modulation are determined by the selfconsistent 
calculation based on the quasiclassical theory.  
That is, we also selfconsistently obtain the radial dependence 
around vortex in the FFLO vortex states, which was not done before.  

On the other hand, the vortex and FFLO nodal plane structures 
in the FFLO state were calculated by the Bogoliubov de-Gennes (BdG) theory 
for a single vortex in a superconductor under a cylindrical symmetry 
situation.~\cite{mizushima2}  
This study clarifies that the topological structure of the pair potential 
plays important roles to determine the electronic structures 
in the FFLO vortex state. 
The pair potential has $2\pi$-phase winding around the vortex line, and 
$\pi$-phase shift at the nodal plane of the FFLO modulation. 
These topologies of the pair potential structure affect the distribution 
of paramagnetic moment and low energy electronic states  
inside the superconducting gap. 
For an example, the paramagnetic moment is enhanced at the vortex core 
and the FFLO nodal plane. 
These structures are related to the bound states 
due to the $\pi$-phase shift of the pair potential.

The other method to microscopically study the vortex state 
is the calculations by the quasiclassical Eilenberger theory. 
In the calculation by the  BdG theory, 
we have to assume a cylindrical symmetry and open 
boundary condition for the system, 
because of the limit of matrix size 
to numerically solve the BdG equation in the present status of 
the computers. 
However, using the Eilenberger theory in the limit 
$k_{\rm F} \xi \ll 1 $  ($k_{\rm F}$ is Fermi wave number and 
$\xi$ is the superconducting coherence length), 
we are free from these restrictions. 
Therefore, 
we can discuss the cases of anisotropic pairing symmetry, 
such as $d$-wave pairing,  
and arbitrary shape Fermi surfaces, 
where we obtain broken cylindrical structures around vortices.  
Since we can consider the system of vortex lattice and periodic 
FFLO modulation by the  periodic boundary condition, 
we can discuss the overlaps between tails of 
the neighbor vortex cores or FFLO nodal planes. 
These calculations for the periodic systems make us possible 
to estimate the resonance line shapes in the NMR experiments 
and form factors in the neutron scattering experiment.  

The purpose of this paper is to investigate the 3D  
structures  of the vortex and the FFLO nodal plane 
in the FFLO vortex state by the quasiclassical theory.  
We calculate the spatial structures of pair potentials, 
paramagnetic moments, internal magnetic fields and 
electronic states in the vortex lattice state 
under the given period of the FFLO modulation. 
From the results, 
we confirm the fundamental properties of the FFLO vortex states 
obtained by the BdG theory, and clarify the further details 
of the FFLO vortex states in the system of the vortex lattice and 
periodic FFLO modulation. 
Since the superconducting pairing symmetry of ${\rm CeCoIn_5}$ is 
suggested to be a $d$-wave with line nodes, we calculate both cases 
of the isotropic $s$-wave pairing and the anisotropic $d$-wave paring 
comparatively, in order to study the contribution of the pairing symmetry 
to the FFLO vortex state.\cite{ikeda,Vorontsov}  
In the calculation, we use the Q2D Fermi surface and a magnetic field is 
applied to the $ab$ plane, as is done for ${\rm CeCoIn_5}$. 
As possible experimental methods to detect the FFLO modulation, 
we discuss the NMR resonance line shape~\cite{kakuyanagi} and 
the neutron scattering experiments. 
We show that 
the spatial structure of the paramagnetic moment reflects the 
NMR resonance line shape in the NMR experiment 
observing Knight shift under an applied magnetic field.

After introducing the formulation for the calculation of 
the FFLO vortex structure 
by the quasiclassical theory including the paramagnetic effect 
in Sec. \ref{sec:formulation}, 
we investigate the spatial structure of the FFLO vortex states 
in Sec. \ref{sec:FFLO}, 
and electronic states in Sec. \ref{sec:LDOS}. 
Using the calculated structure of the FFLO vortex states, 
we discuss the NMR resonance line shape in Sec. \ref{sec:NMR} 
and the neutron scattering in Sec. \ref{sec:SANS}. 
The last section is devoted to the summary and discussions. 

\section{Quasiclassical theory including paramagnetic effect}
\label{sec:formulation}

In this study for the FFLO vortex state, 
we take account of the paramagnetic depairing effect 
due to the Zeeman splitting term $\mu_{\rm B}B({\bf r})$ 
in addition to the orbital depairing effect 
by the vector potential ${\bf A}({\bf r})$. 
Therefore, Hamiltonian for the spin-singlet pairing superconductor 
is given by 
\begin{eqnarray} && 
{\cal H}-\mu_0 {\cal N} 
=\sum_{\sigma=\uparrow,\downarrow}\int{\rm d}^3 {\bf r}
\ \psi_\sigma^\dagger({\bf r}) 
 K_\sigma({\bf r}) 
\psi_\sigma({\bf r})
\nonumber \\ && \quad 
-\int{\rm d}^3 {\bf r}_1\int{\rm d}^3 {\bf r}_2 \left\{ 
 \Delta({\bf r}_1,{\bf r}_2) \psi_\uparrow^\dagger({\bf r}_1) 
\psi_\downarrow^\dagger({\bf r}_2) \right.
\nonumber \\ &&
\qquad \qquad \qquad \left.
+\Delta^\ast({\bf r}_1,{\bf r}_2) \psi_\downarrow({\bf r}_2) 
\psi_\uparrow({\bf r}_1) 
\right\} 
\end{eqnarray} 
with 
\begin{eqnarray}
K_\sigma({\bf r})= 
\frac{\hbar^2}{2m}\left(\frac{\nabla}{\rm i}
+\frac{2 \pi}{\phi_0}{\bf A}\right)^2 
+\sigma\mu_{\rm B}B({\bf r}) -\mu_0 ,
\end{eqnarray} 
the flux quantum $\phi_0$, and   
$\sigma=\pm 1$ for up/down spin electrons. 
In the following, length, temperature, Fermi velocity, 
magnetic field and vector potential are, respectively, scaled by 
$R_0$, $T_c$, $\bar{v}_{\rm F}$, $B_0$ and $B_0 R_0$. 
Here, $R_0=\hbar \bar{v}_{\rm F}/2 \pi k_{\rm B} T_{\rm c}$, 
$B_0=\hbar c /2|e|R_0^2$, and $\bar{v}_{\rm F}=\langle v_{\rm F}^2 
\rangle_{\bf k}^{1/2}$ 
is an averaged Fermi velocity on the Fermi surface. 
$\langle \cdots \rangle_{\bf k}$ indicates the Fermi surface average. 
Energy $E$, pair potential $\Delta$ and Matsubara frequency $\omega_l$ 
are scaled by $\pi k_{\rm B} T_{\rm c}$. 

Using the quasiclassical Green's functions 
$g( \omega_l +{\rm i} \tilde{\mu} B, {\bf k},{\bf r})$, 
$f( \omega_l +{\rm i} \tilde{\mu} B, {\bf k},{\bf r})$,  
$f^\dagger( \omega_l +{\rm i} \tilde{\mu} B, {\bf k},{\bf r})$, 
Eilenberger equations are given 
by~\cite{tachiki,klein,klein87,ichiokaS,ichiokaD,ichiokaMgB2}  
\begin{eqnarray} && 
\left\{ \omega_l + {\rm i}\tilde{\mu} B 
        + {\bf v} \cdot 
           \left(\nabla + {\rm i}{\bf A} \right) \right\} f 
= \Delta({\bf r},{\bf k}) g  , 
\nonumber \\  && 
\left\{ \omega_l + {\rm i}\tilde{\mu} B 
        - {\bf v} \cdot 
           \left(\nabla - {\rm i}{\bf A} \right) \right\} f^\dagger  
= \Delta^\ast({\bf r},{\bf k}) g ,  
\nonumber \\ &&
{\bf v} \cdot {\nabla} g 
= \Delta^\ast({\bf r},{\bf k}) f  - \Delta({\bf r},{\bf k}) f^\dagger ,   
\label{eq:Eil} 
\end{eqnarray} 
where 
$g=(1-ff^\dagger)^{1/2}$, ${\rm Re}g >0$, 
$\Delta({\bf r},{\bf k})=\Delta({\bf r})\phi({\bf k})$, and 
$\tilde{\mu}=\mu_{\rm B} B_0/\pi k_{\rm B}T_{\rm c}$. 
In Eq. (\ref{eq:Eil}), ${\bf r}=({\bf r}_1+{\bf r}_2)/2$ is the 
center of mass coordinate of the Cooper pair, 
and ${\bf k}=(k_a,k_b,k_c)$ is a 
relative momentum of the Cooper pair. 
We set the pairing function $\phi({\bf k})=1$ in the $s$-wave pairing,  
and $\phi({\bf k})=\phi_{x^2-y^2}({\bf k})=
\sqrt{2}(k_a^2-k_b^2)/(k_a^2+k_b^2)$ or 
$\phi({\bf k})=\phi_{xy}({\bf k})=2\sqrt{2} k_a k_b /(k_a^2+k_b^2)$  
in the $d$-wave pairing. 
The vector potential is given by 
${\bf A}=\frac{1}{2} \bar{\bf B}\times{\bf r}+{\bf a}$ 
in the symmetric gauge, with an average flux density 
$\bar{\bf B}$. 
The internal field is obtained as 
${\bf B}({\bf r})=\bar{\bf B}+\nabla\times{\bf a}$. 

The pair potential is selfconsistently calculated by 
\begin{eqnarray}
\Delta({\bf r})
= g_0N_0 T \sum_{0 \le \omega_l \le \omega_{\rm cut}} 
 \left\langle \phi^\ast({\bf k}) \left( 
    f +{f^\dagger}^\ast \right) \right\rangle_{\bf k} 
\label{eq:scD} 
\end{eqnarray} 
with 
$(g_0N_0)^{-1}=  \ln T +2 T
        \sum_{0 \le \omega_l \le \omega_{\rm cut}}\omega_l^{-1} $. 
We use $\omega_{\rm cut}=20 k_{\rm B}T_{\rm c}$.
The vector potential is selfconsistently determined by 
the paramagnetic moment and the supercurrent as 
\begin{eqnarray}
\nabla\times  \nabla \times {\bf a}({\bf r}) 
={\bf j}_{\rm s}({\bf r})+\nabla\times {\bf M}_{\rm para}({\bf r})
\equiv{\bf j}({\bf r}), 
\label{eq:rotA}
\end{eqnarray} 
where 
\begin{eqnarray} 
{\bf j}_{\rm s}({\bf r})
=-\frac{2T}{{\tilde{\kappa}}^2}  \sum_{0 \le \omega_l} 
 \left\langle {\bf v}_{\rm F} 
         {\rm Im}\{ g \}  
 \right\rangle_{\bf k} 
\label{eq:scH} 
\end{eqnarray} 
is the supercurrent, and 
\begin{eqnarray}
M_{\rm para}({\bf r})
=M_0 \left( 
\frac{B({\bf r})}{\bar{B}} 
- \frac{2T}{\tilde{\mu} \bar{B} }  
\sum_{0 \le \omega_l}  \left\langle {\rm Im} \left\{ g \right\} 
 \right\rangle_{\bf k}
\right) 
\label{eq:scM} 
\end{eqnarray} 
is the paramagnetic moment in the vortex lattice state. 
Here, the normal state paramagnetic moment 
$M_0 = ({\tilde{\mu}}/{\tilde{\kappa}})^2 \bar{B} $, 
$\tilde{\kappa}=B_0/\pi k_{\rm B}T_{\rm c}\sqrt{8\pi N_0}$,  
$N_0$ is the density of states at the Fermi energy in the normal state. 

When we calculate the electronic states,  
we solve Eq. (\ref{eq:Eil}) with $ i\omega_l \rightarrow E + i \eta$.  
The local density of states (LDOS) is given by 
\begin{eqnarray} 
N_\sigma({\bf r},E)=\langle {\rm Re }
\{ 
g( \omega_l +{\rm i} \sigma\tilde{\mu} B, {\bf k},{\bf r}) 
|_{i\omega_l \rightarrow E + i \eta} \}\rangle_{\bf k} 
\end{eqnarray} 
for each spin component. 
We typically use $\eta=0.01$. 

As a model of Fermi surface in ${\rm CeCoIn_5}$, 
we use a Q2D Fermi surface with rippled cylinder-shape, 
and the Fermi velocity is given by 
${\bf v}_{\rm F}=(v_a,v_b,v_c) \propto 
( \cos\theta, \sin\theta, \tilde{v}_z \sin k_c)$ 
at the Fermi surface 
${\bf k}_{\rm F}=(k_a,k_b,k_c)
\propto(k_{\rm F0}\cos\theta,k_{\rm F0}\sin\theta,k_c)$ 
with $-\pi \le \theta \le \pi$ and $-\pi \le k_c \le \pi$.~\cite{ichiokaMgB2} 
In our calculation we set $\tilde{v}_z =0.5$, so that the anisotropy ratio 
$\gamma=\xi_c/\xi_{ab} \sim  
\langle v_c^2 \rangle_{\bf k}^{1/2} /\langle v_a^2 \rangle_{\bf k}^{1/2} 
\sim 0.5$. 
A magnetic field is applied along the $a$ axis direction in our calculation. 
Thus, the coordinate $(x,y,z)$ for the vortex structure corresponds to 
$(b,c,a)$ of the crystal coordinate. 
In the $d$-wave pairing, 
the case of the pairing function $\phi_{x^2-y^2}$  
($\phi_{xy}$) corresponds to 
the configuration when $\bar{\bf B}$ is applied to the anti-nodal direction 
(the nodal direction) of the superconducting gap. 

\begin{figure}[tb]
 \includegraphics[width=8.5cm]{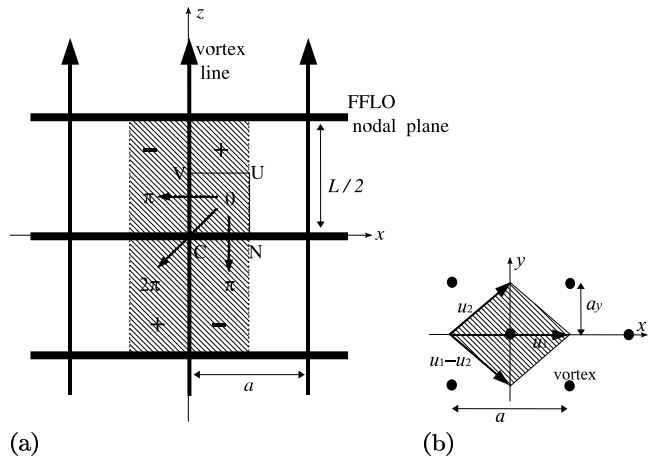} 
  \caption{
Configurations of the vortex lines and the FFLO nodal planes are  
schematically presented in the $xz$ plane including vortex lines (a) 
and in the $xy$ plane (b). 
The vortex distance is $a$ in the $x$ direction, and the distance between  
the FFLO nodal planes is $L/2$. 
The hatched region indicates the unit cell. 
In (a), along the trajectories presented by ``$0 \longrightarrow \pi$'', 
the pair potential changes the sign ($+ \rightarrow -$) 
across the vortex line or across the FFLO nodal plane, 
due to the $\pi$-phase shift of the pair potential.   
Along the trajectory presented by ``$0 \longrightarrow 2\pi$'', 
the sign of the the pair potential does not change ($+ \rightarrow +$) 
across the intersection point of the vortex line and the FFLO nodal plane, 
since the phase shift is $2\pi$. 
In (b), $\bullet$ indicates the vortex center. 
${\bf u}_1-{\bf u}_2$ and ${\bf u}_2$ are unit vectors of the vortex 
lattice.  
}
  \label{fig:unit}
\end{figure}

We solve Eq. (\ref{eq:Eil}) and Eqs. (\ref{eq:scD})-(\ref{eq:scM}) 
alternately, and obtain selfconsistent solutions 
as in previous works,~\cite{ichiokaS,ichiokaD,ichiokaMgB2} 
by fixing a unit cell of the vortex lattice and 
a period $L$ of the FFLO modulation.  
The unit cell of the vortex lattice is given by 
${\bf r}=w_1({\bf u}_1-{\bf u}_2)+w_2{\bf u}_2+w_3{\bf u}_3$ with 
$-0.5 \le w_i \le 0.5$ ($i$=1, 2, 3), ${\bf u}_1=(a,0,0)$,  
${\bf u}_2=(a/2,a_y,0)$ and ${\bf u}_3=(0,0,L)$. 
Reflecting anisotropy ratio $\gamma$, we set 
$a_y/a=\sqrt{3}\gamma/2$ with $\gamma =0.5$. 
For the FFLO modulation, we assume 
$\Delta(x,y,z)=\Delta(x,y,z+L)$ and 
$\Delta(x,y,z)=-\Delta(x,y,-z)$. 
Then, $\Delta({\bf r})=0$ at the FFLO nodal planes  
$z=0$, and $\pm 0.5L$. 
These configurations of the FFLO vortex structure are schematically shown 
in Fig. \ref{fig:unit}, which show the unit cell in the $xz$ plane 
including vortex lines, and in the $xy$ plane.

We divide $w_i$ to $N_i$-mesh points in our numerical calculation, 
and obtain the quasiclassical Green's functions, $\Delta({\bf r})$, 
$M_{\rm para}({\bf r})$ and ${\bf j}({\bf r})$ at each mesh point 
in the 3D space.  
Typically we set $N_1=N_2=N_3=31$. 
In the selfconsistent calculation of ${\bf a}$, 
we solve Eq. (\ref{eq:rotA}) in the Fourier space 
${\bf q}_{m_1,m_2,m_3}$,  
taking account of the current conservation 
$\nabla \cdot {\bf j}({\bf r})=0$,  
so that the average flux density per unit cell of the vortex lattice 
is kept constant.  
The wave number ${\bf q}$ is discretized as 
\begin{equation} 
{\bf q}_{m_1,m_2,m_3}=m_1{\bf q}_1+m_2{\bf q}_1+m_3{\bf q}_3 
\label{eq:q123}
\end{equation} 
with integers $m_i$ ($i=1,2,3$), 
where ${\bf q}_1=(2 \pi/a, -\pi/a_y,0)$,  
${\bf q}_2=(2 \pi/a, \pi/a_y,0)$, and ${\bf q}_3=(0,0,2 \pi/L)$. 
The lattice momentum is defined as 
${\bf G}({\bf q}_{m_1,m_2,m_3})=(G_x,G_y,G_z)$ with 
$G_x=[N_1 \sin(2 \pi m_1/N_1)+N_2 \sin(2 \pi m_2/N_2)]/a$,  
$G_y=[-N_1 \sin(2 \pi m_1/N_1)
      +N_2 \sin(2 \pi m_2/N_2)]/2 a_y$,  and 
$G_z=N_3 \sin(2 \pi m_3/N_3) /L$.  
We obtain the Fourier component of ${\bf a}({\bf r})$ as    
${\bf a}({\bf q})={\bf j}'({\bf q})/|{\bf G}|^2$, where  
${\bf j}'({\bf q})={\bf j}({\bf q}) 
-{\bf G}\left( {\bf G}\cdot{\bf j}({\bf q}) \right)/|{\bf G}|^2 $  
ensuring the current conservation $\nabla \cdot {\bf j}'({\bf r})=0$, 
and ${\bf j}({\bf q})$ is the Fourier component of 
${\bf j}({\bf r})$ in Eq. (\ref{eq:rotA}).\cite{klein87}    
The final selfconsistent solution satisfies 
$\nabla \cdot {\bf j}({\bf r})=0$. 
When we solve Eq. ({\ref{eq:Eil}) by the explosion 
method,~\cite{klein87,ichiokaS,ichiokaD,ichiokaMgB2} 
we estimate $\Delta({\bf r})$ and ${\bf A}({\bf r})$ at arbitrary 
positions by the interpolation from their values at the mesh points, 
and by the periodic boundary condition of the unit cell including the 
phase factor due to the magnetic field. 

\section{FFLO vortex structure}
\label{sec:FFLO}

\begin{figure*}[tb]
\includegraphics[width=16cm]{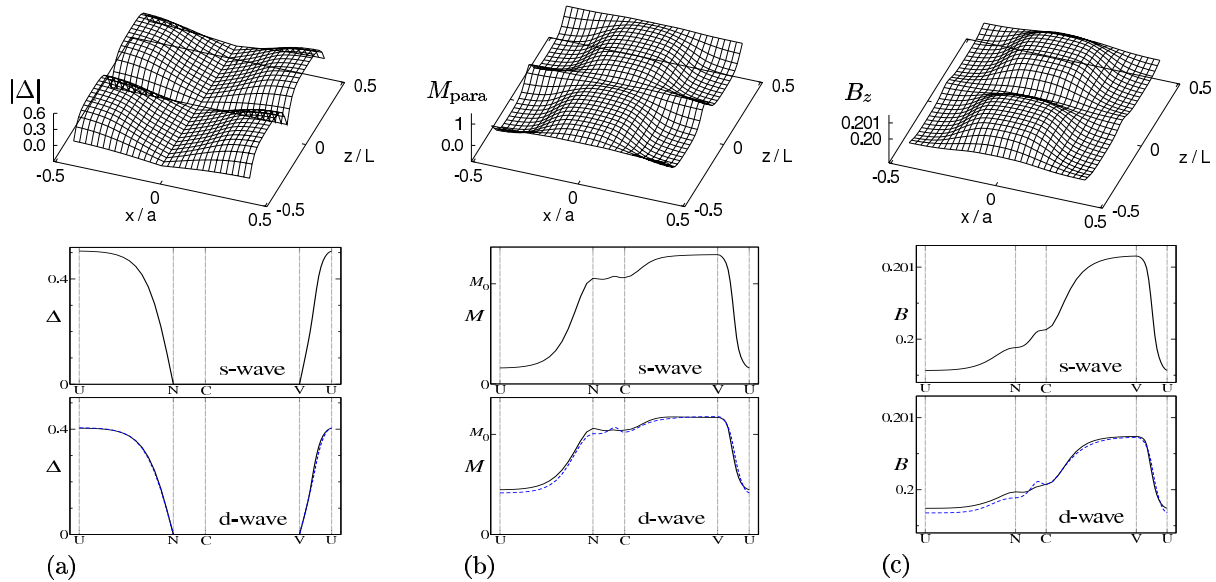} 
  \caption{
(color online) 
 Spatial structure of the FFLO vortex state in the $xz$ plane 
at $\bar{B}=0.2B_0$, $T=0.1T_{\rm c}$  and $L=50R_0$. 
(a) Amplitude of the pair potential $|\Delta({\bf r})|$. 
(b) Paramagnetic moment $M_{\rm para}({\bf r})$. 
(c) Internal magnetic field $B_z({\bf r})$. 
The upper panels show the spatial variation within a unit cell 
[hatched region in Fig. \ref{fig:unit}(a)] for the $s$-wave pairing. 
The middle panels present the profiles  
along the path UNCVU shown in Fig. \ref{fig:unit}(a)   
for the $s$-wave pairing. 
The lower panels are the same as the middle panels, but 
for the $d$-wave pairing. 
The solid (dashed) lines are for the case when the applied field 
is applied to the anti-nodal (nodal) direction. 
}
  \label{fig:DMB}
\end{figure*}

We consider the case of large GL parameter   
$\tilde{\kappa}=20$ and large paramagnetic 
contribution $\tilde{\mu}=1.7$. 
In these parameters, 
the upper critical field is suppressed 
to $H_{c2}(T=0.1T_c) \sim 0.22 B_0$,  
as a first order transition. 
We discuss the FFLO vortex state 
when we set $L=50R_0$ or $70R_0$ at $T=0.1T_c$ 
and $\bar{B}=0.2 B_0$, giving $M_0=0.00139 B_0$ and 
the vortex lattice constant $a=8.5R_0$.

In the upper panels of Fig. \ref{fig:DMB}, 
we show the spatial structure of the FFLO vortex state
within a unit cell in the slice of the $xz$ plane, i.e., the hatched region 
shown in Fig. \ref{fig:unit}(a). 
In the middle and lower panels of Fig. \ref{fig:DMB}, 
the profiles of the spatial structure are presented 
along the path UNCVU shown in Fig. \ref{fig:unit}(a). 
The point C ($x=y=z=0$) is the intersection point of 
a vortex and a nodal plane. 
The point V ($x=y=0$ and $z=L/4$) is at the vortex center and 
far from the FFLO nodal plane. 
The point N ($x=a/2$, $y=z=0$) is at the FFLO nodal plane 
and outside of the vortex.   
The point U ($x=a/2$, $y=0$, $z=L/4$) is 
far from both the vortex and the FFLO nodal plane. 
In Fig. \ref{fig:DMB}, 
in addition to the $s$-wave pairing case in the middle panels,  
we also present the $d$-wave pairing cases in the lower panels  
in order to show that the qualitative features of 
the FFLO vortex structure are independent of the pairing symmetry 
and the field orientation within the $ab$ plane. 

Figure \ref{fig:DMB}(a) shows the amplitude of the order parameter. 
In the upper panel, we see that  $|\Delta({\bf r})|$ 
is suppressed near the vortex center 
at $x=y=0$ and the FFLO nodal plane at $z=0$, $\pm 0.5L$.
Far from the FFLO nodal plane such as $z=0.25L$, 
$|\Delta({\bf r})|$ shows a typical profile of the conventional vortex. 
When we cross the vortex line or the FFLO nodal plane, 
the sign of $\Delta({\bf r})$ changes due to the $\pi$-phase shift 
of the pair potential as schematically shown in Fig. \ref{fig:unit}(a). 
In the profile of $|\Delta({\bf r})|$ presented in the 
middle and the lower panels of Fig. \ref{fig:DMB}(a), 
$|\Delta({\bf r})|=0$ along the FFLO nodal plane NC and 
along the vortex line CV. 
While amplitude $|\Delta({\bf r})|$ in the $s$-wave pairing [middle panel] 
is larger than that in the $d$-wave pairing [lower panel], 
this is due to the difference of the ratio $\Delta(H=0)/T_c$ 
depending on the pairing symmetry.  

Correspondingly, paramagnetic moment $M_{\rm para}({\bf r})/M_0$ 
is presented in Fig. \ref{fig:DMB}(b). 
As is well-known as the Knight shift, 
the paramagnetic moment is suppressed 
in the spin-singlet pairing superconducting state.  
In the figures, we see that $M_{\rm para}({\bf r})$ 
is suppressed outside of the vortex and far from the FFLO nodal plane, 
as expected.  
However, $M_{\rm para}({\bf r})$ is enhanced at the vortex core 
or at the FFLO nodal plane, 
both in the $s$-wave and the $d$-wave pairings. 
The reason for these structures of $M_{\rm para}({\bf r})$ is discussed 
later in connection with the LDOS. 
At the FFLO nodal plane $M_{\rm para}({\bf r}) \sim M_0$ 
[path NC in Fig. \ref{fig:DMB}(b)].   
Along the vortex line, 
$M_{\rm para}({\bf r})$ is enhanced more than $M_0$  
far from the FFLO nodal planes 
[position V in Fig. \ref{fig:DMB}(b)]. 

Figure \ref{fig:DMB}(c) presents the $z$-component 
of the internal field, $B_z({\bf r})$. 
Due to the contribution of the enhanced $M_{\rm para}({\bf r})$, 
$B_z({\bf r})$ is enhanced at the FFLO nodal plane 
even outside of the vortex.  
A part of the contributions by $M_{\rm para}({\bf r})$ is compensated 
by the diamagnetic contribution, 
because the average flux density per unit cell of the vortex lattice 
in the $xy$ plane should conserve along the magnetic field direction. 
Therefore, due to the conservation, 
the enhancement of $B_z({\bf r})$ at the FFLO nodal plane 
[path NC in Fig. \ref{fig:DMB}(c)] is smaller, 
compared with the enhancement of $M_{\rm para}({\bf r})$ 
at the FFLO nodal plane [path NC in Fig. \ref{fig:DMB}(b)]. 
While $B_z({\bf r})$ is largely enhanced than $\bar{B}$ 
at the vortex core far from the FFLO nodal plane 
[position V in Fig. \ref{fig:DMB}(c)], 
$B_z({\bf r})$ is not largely enhanced at the vortex core 
in the FFLO nodal plane [position C]. 
Therefore $B_z({\bf r}) \sim \bar{B}=0.2$ at the FFLO nodal plane 
[path NC]. 

We compare the behaviors of the middle panels for the $s$-wave pairing  
and the lower panels for the $d$-wave pairing. 
The profile along the path VU is almost the same as  that 
of the conventional vortex without FFLO modulation. 
Quantitatively, the variable ranges of $M_{\rm para}({\bf r})$ and 
$B_z({\bf r})$ are smaller in the $d$-wave pairing, 
compared with the $s$-wave pairing.  
In the lower panels, we also show the dependence 
on the relative angle of the magnetic field and the gap nodes.  
In any cases of pairings and the magnetic field directions, 
we see qualitatively the same behavior of the FFLO structure, 
i.e., $M_{\rm para}({\bf r})\sim M_0$ and $B_z({\bf r}) \sim \bar{B}$.  
Therefore, 
if we study the dependence on the pairing symmetry and 
the magnetic field direction within the $ab$ plane,    
we need to carefully examine the quantitative difference of 
the FFLO structure in the calculation of the pair potential, 
the paramagnetic moment and the internal magnetic field. 
On the other hand, the dependences on the pairing symmetry
appear more clearly in the structure of the electronic 
states in the vortex states.\cite{ichiokaD}  
Therefore, we discuss the electronic structure 
in the FFLO vortex states in the next section.

\section{Electronic structure in the FFLO vortex state}
\label{sec:LDOS}

\begin{figure}[tb]
\includegraphics[width=4.3cm]{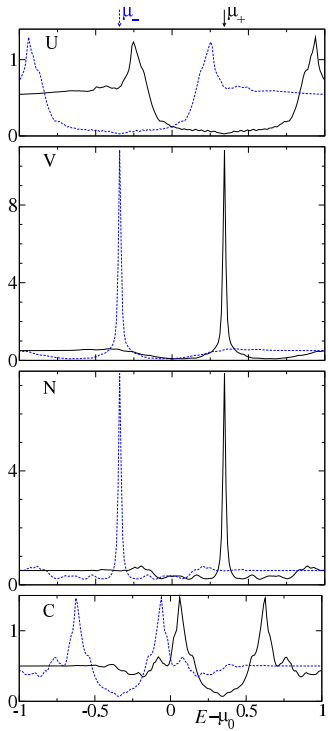} 
  \caption{
(color online) 
Spectrum of the LDOS for up-spin electrons 
$N_\uparrow(E,{\bf r})/N_0$ (solid lines) and 
for down-spin electrons $N_\downarrow(E,{\bf r})/N_0$ (dashed lines) 
in the $s$-wave pairing. 
$\bar{B}=0.2 B_0$, $T=0.1T_{\rm c}$ and $L=50 R_0$. 
From top panel to bottom, we present the LDOS 
at positions U, V, N, and C.  
Their locations are shown in Fig. \ref{fig:unit}(a). 
}
  \label{fig:LDOSs}
\end{figure}
\begin{figure}[tb]
\includegraphics[width=8.5cm]{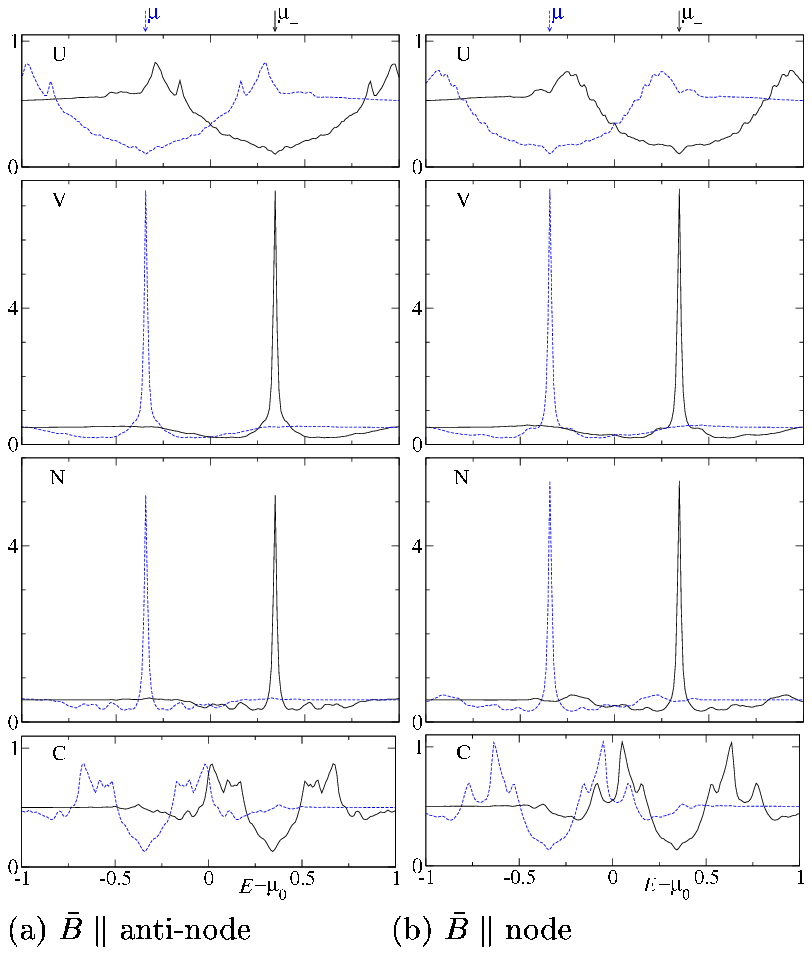} 
  \caption{
(color online) 
The same as Fig. \ref{fig:LDOSs}, but in the $d$-wave pairing. 
(a) $\bar{{\bf B}}$ is applied to the anti-nodal direction. 
(b) $\bar{{\bf B}}$ is applied to the nodal direction. 
}
  \label{fig:LDOSd}
\end{figure}

The LDOS spectrum for up- and down-spin electrons  
are presented at some positions 
in Fig. \ref{fig:LDOSs} for the $s$-wave pairing and 
in Fig. \ref{fig:LDOSd} for the $d$-wave pairing. 
In the quasiclassical theory,  
$N_\sigma(E,{\bf r})$ are symmetric by $E \leftrightarrow -E$ 
in the absence of the paramagnetic effect ($\tilde{\mu}=0$).  
In the presence of the paramagnetic effect, 
the LDOS spectrum for up- (down-) spin electrons is shifted 
to positive (negative) energy by $\tilde{\mu}\bar{B}$ due to the 
Zeeman shift. 
In this case, we have a relation 
$N_\uparrow(E,{\bf r})=N_\downarrow(-E,{\bf r})$ 
within the quasiclassical theory. 

Far from the FFLO nodal plane and outside of vortex,  
as shown in the spectrum of the top panel for the position U 
in Fig. \ref{fig:LDOSs}, 
we see the superconducting full-gap structure of the $s$-wave pairing.  
There, small LDOS also appears at low energies inside the gap  
due to the low energy excitations extending from the vortex cores 
and the FFLO nodal planes at finite magnetic fields. 
Since the LDOS at $E<0$ ($E>0$) are occupied (empty) states, 
there is a relation between the LDOS spectrum 
and local paramagnetic moment as  
\begin{eqnarray} 
M_{\rm para}({\bf r})=-\mu_{\rm B}\int_{-\infty}^0
(N_\uparrow(E,{\bf r}) -N_\downarrow(E,{\bf r}) ){\rm d}E.
\label{eq:M-LDOS}
\end{eqnarray}  
In the full-gap case of the $s$-wave pairing, 
the difference of the occupation number between up- and down-spin electrons 
is small, since the LDOS below the gap are 
occupied similarly in $N_\uparrow(E,{\bf r})$ and $N_\downarrow(E,{\bf r})$. 
This is the reason why $M_{\rm para}({\bf r})$ is suppressed 
at the position U. 
Small but finite $M_{\rm para}({\bf r})$ comes from 
the small LDOS weight of low energy states inside the gap 
in the top panel of Fig. \ref{fig:LDOSs}. 
 
The LDOS spectra at the position V on the vortex center and 
at the position N on the FFLO nodal plane are, respectively, 
presented in the second and the third panels of 
Fig. \ref{fig:LDOSs}.   
In these figures,  
$N_\uparrow(E,{\bf r})$ and $N_\downarrow(E,{\bf r})$, respectively, 
have a sharp peak at $E=\mu_+$ and $E=\mu_-$,  
with  $\mu_\pm \equiv \mu_0 \pm \tilde{\mu}\bar{B}$. 
These peaks are related to the topological structure 
of the pair potential, as schematically shown in Fig.  \ref{fig:unit}. 
Since a vortex has phase winding $2\pi$, 
along the trajectory through the vortex center,  
$\Delta({\bf r})$ changes the sign by the $\pi$-phase shift  
across the vortex center.  
Also at the trajectory through the FFLO nodal plane, 
$\Delta({\bf r})$ changes the sign across the nodal plane. 
The bound states appear when the pair potential has 
the $\pi$-phase shift, and form the ``zero-energy peak''.  
This peak is  shifted to $E=\mu_+$ or $E=\mu_-$ 
due to the Zeeman effect.\cite{mizushima2,Takahashi}  
Since the peak of the LDOS spectrum for up-spin electrons 
is an empty state ($E>0$) and the peak of the LDOS for down-spin electrons 
is an occupied state ($E<0$), 
$M_{\rm para}({\bf r})$ becomes large at these positions, 
from the relation in Eq. ({\ref{eq:M-LDOS}).
 
On the other hand, 
along the trajectory through the intersection point of 
a vortex and a nodal plane, $\Delta({\bf r})$ does not change the sign, 
because the phase shift is $2 \pi$ by summing $\pi$ due to vortex and 
$\pi$ due to the nodal plane, 
as schematically shown in Fig.  \ref{fig:unit}. 
Thus, the sharp peaks do not appear at $E=\mu_\pm$  
as seen from the bottom panel of Fig. \ref{fig:LDOSs}.  
Instead,  $N_\uparrow(E,{\bf r})$  has two broad peaks at finite energies  
shifted upper or lower from $\mu_+$. 
In this situation, 
$M_{\rm para}({\bf r})$ is still large at position C, 
as in positions V and N,  
since the LDOS in both peaks are empty ($E>0)$ in $N_\uparrow(E,{\bf r})$, 
and occupied ($E<0$) in $N_\downarrow(E,{\bf r})$.

The differences of the $s$-wave and the $d$-wave pairings appear    
in the LDOS spectrum of the superconducting gap. 
At a zero field, the LDOS has the full-gap structure in the $s$-wave 
pairing, while it has a V-shape gap in the $d$-wave pairing with line nodes. 
Under the magnetic field, these features of the gap structure are 
smeared by low energy excitations under the magnetic field, but 
can be seen far from the vortex and the FFLO nodal plane,  
as is shown in top panels [position U] 
in Figs. \ref{fig:LDOSs} and \ref{fig:LDOSd}. 
In the $d$-wave pairing, the LDOS spectrum has 
larger weight of low energy states within the gap 
[top panels in Fig. \ref{fig:LDOSd}], and $M_{\rm para}({\bf r})$ is larger 
[position U in Fig. \ref{fig:DMB}(b)], 
compared with the $s$-wave pairing. 
From Fig. \ref{fig:LDOSd}, we see that in the $d$-wave pairing 
the LDOS structures at the vortex line and at the FFLO nodal plane 
are qualitatively the same as in the $s$-wave pairing. 
The sharp peak structure at $E=\mu_\pm$ appears in the LDOS 
at the vortex line [second panels] and 
at the FFLO nodal plane [third panels]. 
And the peak is split to two broad peaks, 
and shifted to upper or lower from $\mu_\pm$ 
at the intersection point of the vortex line and the FFLO nodal 
plane [bottom panels for the position C].
Figure \ref{fig:LDOSd} presents two cases of 
the magnetic field orientation, which show that 
the characteristic peak structures of the LDOS 
in the FFLO vortex structure do not significantly depend 
on the pairing symmetry and the relative angle of the magnetic 
field orientation and the gap-node direction.   

\begin{figure}[tb]
\includegraphics[width=9.0cm]{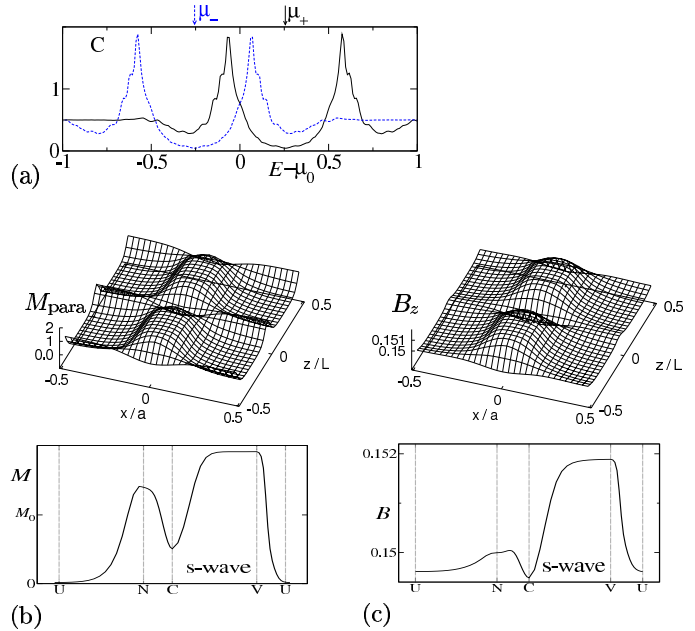} 
  \caption{
(color online) 
FFLO vortex structure at $\bar{B}=0.15B_0$, $T=0.1T_{\rm c}$ 
and $L=50R_0$ in the $s$-wave pairing. 
(a) Spectrum of the LDOS for up-spin electrons 
$N_\uparrow(E,{\bf r})/N_0$ (solid lines) and 
for down-spin electrons $N_\downarrow(E,{\bf r})/N_0$ (dashed lines) 
at the position C. 
(b) Paramagnetic moment $M_{\rm para}({\bf r})$. 
(c) Internal field $B_z({\bf r})$. 
The upper panels of (b) and (c) show 
the spatial variation within a unit cell. 
The lower panels present the profile 
along the path UNCVU.  
}
  \label{fig:h0.15}
\end{figure}

In order to understand the relation between the LDOS and 
the paramagnetic moment, it is instructive to see the lower 
magnetic field case at $\bar{B}=0.15B_0$ 
shown in Fig. \ref{fig:h0.15}. 
Here, $a=9.8R_0$ and $M_0=0.00105B_0$. 
Figure \ref{fig:h0.15}(a) shows the LDOS at the position C. 
Compared with the LDOS in Fig. \ref{fig:LDOSs}, 
the Zeeman shift $\tilde{\mu}\bar{B}$ of the spectrum is smaller.  
In this case, one of two broad peaks is occupied ($E<0$) and the other 
is empty ($E>0$) both for $N_\uparrow(E,{\bf r})$ and 
$N_\downarrow(E,{\bf r})$. 
Therefore, $M_{\rm para}({\bf r})$ is suppressed at the position C, 
i.e., the intersection point of the vortex line and the FFLO nodal plane, 
as shown in Fig. \ref{fig:h0.15}(b), 
while $M_{\rm para}({\bf r})$ is larger at the vortex line (position V) 
and at the FFLO nodal plane (position N). 
This structure corresponds to that obtained 
in the calculation by the BdG theory.~\cite{mizushima2}  
Also in the spatial structure of the internal field, 
$B_z({\bf r})$ is suppressed at the position C 
as shown in Fig.  \ref{fig:h0.15}(c) 
by the contribution from the spatial structure of $M_{\rm para}({\bf r})$.

\begin{figure}[tb]
\includegraphics[width=5.5cm]{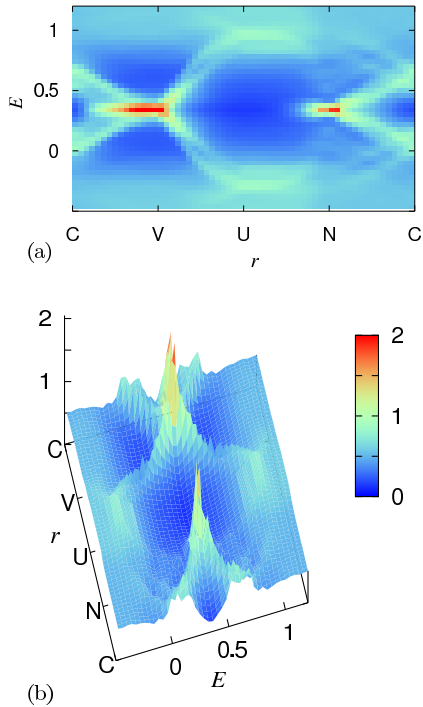} 
  \caption{
(color) 
Spectral evolution of 
the LDOS for up-spin electrons 
along the path CVUNC shown in Fig. \ref{fig:unit}(a). 
Density plot (a) and stereographic view (b) of 
$N_\uparrow(E,{\bf r})/N_0$ 
are presented in the $d$-wave pairing, 
when magnetic field $\bar{B}=0.2 B_0$ is applied 
to the anti-nodal direction. 
$T=0.1T_{\rm c}$ and $L=50 R_0$. 
}
  \label{fig:SPC}
\end{figure}

In Fig. \ref{fig:SPC}, we show the spectral evolution of 
$N_\uparrow(E,{\bf r})/N_0$
around vortex and the FFLO nodal plane,  
when the pairing is $d$-wave and magnetic field is applied 
to the anti-nodal direction within $ab$ plain. 
We obtain similar spectral evolution also in the $s$-wave pairing.  
This spectral evolution is compared with that 
obtained by BdG theory in the $s$-wave pairing [see Fig. 3 
in Ref. \onlinecite{mizushima2}]. 
The spectral structure is qualitatively good accordance with that. 
In the path VU, we see 
a typical spectral evolution of Caroli-de Gennes-Matricon states
around the conventional vortex 
core.~\cite{Takahashi,Caroli,CdGM,Gygi,Hayashi} 
With aparting from the vortex center, 
vortex core states have higher angular momentum, and the energy 
levels are shifted to higher energy.
Along the vortex line CV, 
moving from the nodal plane, 
two peak states at C are merged to the conventional vortex core states 
with $E=\mu_+$ at V.  
Also along the FFLO nodal plane NC, 
two peak states at the vortex center C 
are merged to the bound states with  $E=\mu_+$ 
at the midpoint N of two vortices.     
The calculation for the midpoint was not done 
in the previous BdG calculation assuming cylindrical 
symmetry.~\cite{mizushima2}

\section{NMR spectrum} 
\label{sec:NMR}

\begin{figure}[tb]
\includegraphics[width=8.5cm]{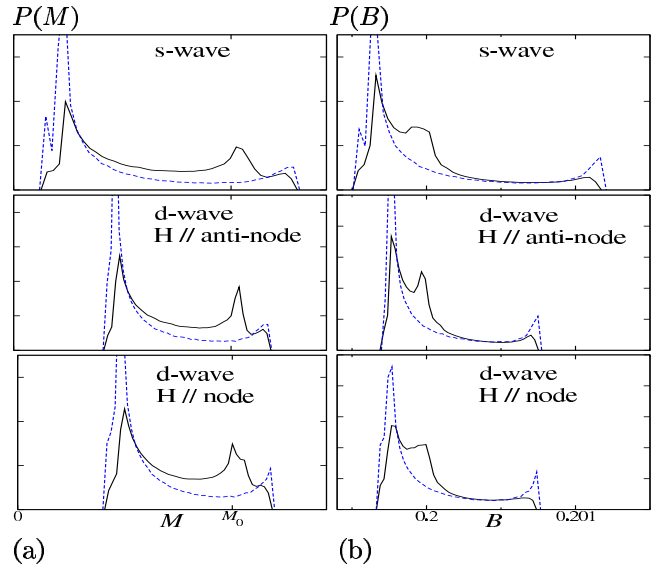} 
  \caption{
(color online) 
(a) Distribution function of the paramagnetic moment. 
We show $P(M)$ as a function of $M_{\rm para}$.  
(b) Distribution function of the internal magnetic field. 
We show $P(B)$ as a function of $B_z$.  
The solid (dashed) lines are for the vortex state with (without) 
the FFLO modulation. 
$\bar{B}=0.2 B_0$, $T=0.1T_c$ and $L=50R_0$.  
The top panels are for the $s$-wave pairing. 
The middle (bottom) panels are for the $d$-wave pairing 
when the magnetic field is applied to the anti-nodal 
(nodal) directions. 
The heights of $P(M)$ and $P(B)$ are scaled so that 
$\int P(M){\rm d}M=\int P(B){\rm d}B=1$. 
  }
  \label{fig:KS}
\end{figure}

In the NMR experiment, resonance frequency spectrum of the nuclear spin 
resonance is determined by the internal magnetic field 
and the hyperfine coupling to the spin of the conduction electrons. 
Therefore, in a simple consideration, the effective field 
for the nuclear spin is given by 
$B_{\rm eff}({\bf r})=B({\bf r})+ A_{\rm hf}M_{\rm para}({\bf r})$, 
where $A_{\rm hf}$ is a hyperfine coupling constant depending 
on species of the nuclear spins. 
The resonance line shape of NMR is given by 
\begin{eqnarray} 
P(\omega)=\int \delta(\omega-B_{\rm eff}({\bf r}) ){\rm d}{\bf r}, 
\end{eqnarray} 
i.e., the intensity at each resonance frequency $\omega$ comes from 
the volume satisfying $\omega=B_{\rm eff}({\bf r})$ in a unit cell. 
When the contribution of the hyperfine coupling is dominant,  
the NMR signal selectively detects $M_{\rm para}({\bf r})$. 
This is the experiment observing the Knight shift 
in superconductors.~\cite{kohori,tou} 
As the NMR spectrum of the Knight shift, we calculate the 
distribution function 
$P(M)=\int \delta(M-M_{\rm para}({\bf r}) ){\rm d}{\bf r}$
from the spatial structure of the paramagnetic moment 
$M_{\rm para}({\bf r})$ shown in Fig. \ref{fig:DMB}(b). 
We show $P(M)$ in Fig. \ref{fig:KS}(a). 
On the other hand, in the case of negligible hyperfine coupling, 
the NMR signal is determined by the internal magnetic field distribution. 
This resonance line shape is called ``Redfield pattern'' 
of the vortex lattice.
In Fig. \ref{fig:KS}(b), we show the distribution function 
$P(B)=\int \delta(B-B_z({\bf r}) ){\rm d}{\bf r}$ 
calculated from the internal field $B_z({\bf r})$. 
In Fig. \ref{fig:KS}, the $s$-wave pairing case and 
two cases of the field orientations in the $d$-wave pairing
are displayed.

First we discuss the line shape of the distribution function $P(M)$.  
The spectrum of $P(M)$ in the conventional vortex state 
without the FFLO modulation is shown 
by dashed lines in Fig. \ref{fig:KS}(a).  
There, the peak of $P(M)$ comes from the signal 
from the outside of the vortex core. 
The temperature dependence of this peak position $M$ 
gives Knight shift observed in the NMR experiments.~\cite{kohori,tou} 
In our calculation, we confirm that $M$ is decreased from $M_0$ 
on lowering temperature below $T_c$.  
The spectrum of $P(M)$ has a tail toward larger $M$ 
by the vortex core contribution with large $M_{\rm para}({\bf r})$. 
The vortex core contribution is a 1D structure, 
their volume contribution is small in the spectrum, 
compared with the peak intensity due to the large volume contribution 
by the outside of the vortex. 
While $P(M)$ is slightly enhanced at the upper edge of $M$, 
this is the feature of the large paramagnetic effect. 
At the higher field near $H_{c2}$, the vortex core radius is 
expanded by the contribution of $M_{\rm para}({\bf r})$ 
enhanced at the vortex core, and the profile of 
$M_{\rm para}({\bf r})$ becomes flatter 
around the expanded vortex core region. 
These contributions slightly enhance $P(M)$ at the upper edge. 
This enhancement is not seen when the paramagnetic effect is not large, 
such as at lower fields. 

In the vortex state with the FFLO modulation, 
the line shape $P(M)$ becomes double peak structure, 
as presented by solid lines in Fig. \ref{fig:KS}(a). 
The height of the main peak decreases,   
and there appears a new peak coming from the FFLO nodal plane 
near $M_{\rm para}\sim M_0$. 
This is because a part of contribution from the outside region of the vortex 
is shifted to the contribution of the FFLO nodal plane. 
The contribution from the two dimensional structure of the FFLO nodal plane 
appears in $P(M)$ more clearly than that of the 1D structure 
of the vortex line. 

Second, we discuss 
the distribution function $P(B)$ of the internal magnetic field 
presented in Fig. \ref{fig:KS}(b). 
There, in the absence of the FFLO modulation (dashed lines), 
the Redfield pattern $P(B)$ has sharp peak corresponding 
the saddle point of the internal field distribution. 
The tail to higher $B$ comes from the increasing magnetic 
field around the vortex core. 
The small enhancement of $P(B)$ at the upper edge of $B$ 
is also due to the contribution from the flat profile of 
$M_{\rm para}({\bf r})$ around the expanded vortex core, 
mentioned above. 
In the presence of the FFLO modulation (solid lines), 
the height of the original peak is decreased, 
and a new peak appears at $B \sim \bar{B}=0.2B_0$ 
as the contribution of the FFLO nodal plane. 
As a feature of $P(B)$, 
the new peak appears near the main peak, 
compared with the line shape of $P(M)$. 

In the $d$-wave pairing seen in lower two panels in Fig. \ref{fig:KS}, 
while the ranges for the variation of $M$ and $B$ are narrower, 
the double peak structures of the FFLO vortex state are seen 
as in the $s$-wave pairing. 
We also see that  
the shapes of $P(M)$ and $P(B)$ are not significantly affected 
by the orientation of the applied magnetic field within the $ab$ plane 
in the line node case of the $d$-wave pairing.

\begin{figure}[tb]
\includegraphics[width=8.5cm]{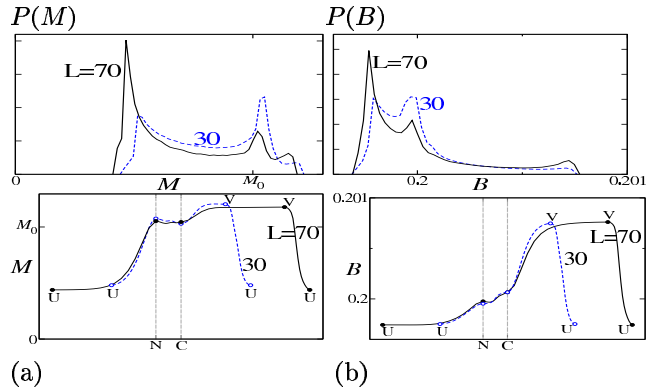} 
  \caption{
(color online) 
Structure of paramagnetic moment (a) and 
internal magnetic field (b) 
when $L/R_0=70$ (solid lines) and 30 (dashed lines) 
in the $d$-wave pairing 
when the magnetic field is applied to the anti-nodal direction.  
$\bar{B}=0.2 B_0$ and $T/T_c=0.1$. 
The upper panels show the distribution function 
$P(M)$ or $P(B)$. 
The lower panels show the profile of $M_{\rm para}({\bf r})$
or $B_z({\bf r})$ along the path UNCVU. 
 }
  \label{fig:KSL}
\end{figure}

\begin{figure}[tb]
\includegraphics[width=8.5cm]{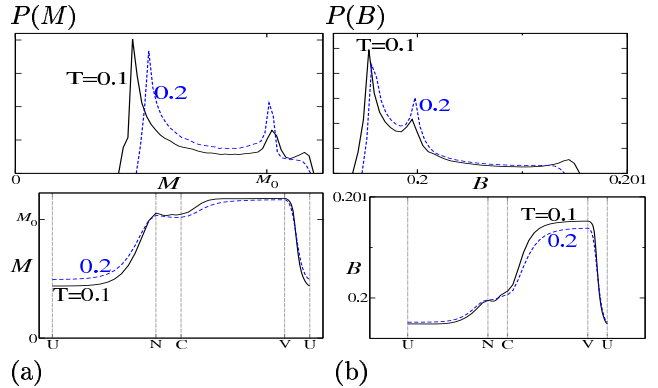} 
  \caption{
(color online) 
Structure of paramagnetic moment (a) and internal magnetic field (b) 
at $T/T_c=$0.1 (solid lines) and 0.2 (dashed lines) 
in the $d$-wave pairing 
when the magnetic field is applied to the anti-nodal direction.   
$\bar{B}=0.2 B_0$ and $L=70 R_0$. 
The upper panels show the distribution function 
$P(M)$ or $P(B)$. 
The lower panels show the profile of $M_{\rm para}({\bf r})$ 
or $B_z({\bf r})$ along the path UNCVU. 
 }
  \label{fig:KST}
\end{figure}

Next, we discuss how the spectra of $P(M)$ and $P(B)$ 
depend on the period $L$ of the FFLO modulation, and on the temperature $T$. 
The $L$-dependences of $P(M)$ and $P(B)$ are shown 
in the upper panels of Fig. \ref{fig:KSL}, 
with the profiles of $M_{\rm para}({\bf r})$ and $B_z({\bf r})$ 
in the lower panels. 
With decreasing $L$, 
the new peak of the FFLO nodal plane is enhanced both in $P(M)$ and $P(B)$, 
since the relative weight of volume contribution 
by the FFLO nodal plane increases in the distribution functions. 
As is seen from the profile along the path UN 
in the lower panel of  Fig.  \ref{fig:KSL}, 
the effective width of FFLO nodal plane where $M_{\rm para}({\bf r})$  
is enhanced near the FFLO nodal plane [position N] 
does not significantly change. 
Instead, the width of the region [around position U] 
with low $M_{\rm para}({\bf r})$ and $B_z({\bf r})$,  
outside of the effective FFLO nodal plane width,   
becomes shorter, when $L$ decreases.  

The temperature dependences of $P(M)$ and $P(B)$ are shown 
in the upper panels of Fig. \ref{fig:KST}, 
with the profiles of $M_{\rm para}({\bf r})$ and $B_z({\bf r})$ 
in the lower panels. 
In the spectra of $P(M)$ and $P(B)$, 
the height of the new peak due to the FFLO nodal plane becomes 
higher with increasing $T$. 
This is because the effective width of FFLO nodal plane 
becomes larger at higher $T$, 
since slope of $M_{\rm para}({\bf r})$ and $B_z({\bf r})$ 
from the FFLO nodal plane decreases 
as shown in the lower panel of Fig. \ref{fig:KST}   
[see slope around the position N in the path UN].  
This enhancement of the new peak of the FFLO nodal plane at higher $T$  
comes from the increase of the effective coherence length. 
On the other hand, the slight enhancement of $P(M)$ and $P(B)$ at the 
upper edge, which appears even in the absence of the FFLO modulation, 
is smeared at higher $T$.  
This difference of the $T$-dependence can be used to 
distinguish the origin of the new peak observed in experiments. 
If the new peak becomes eminent at higher $T$, the origin of the new peak 
is the FFLO nodal plane. 
If the new peak becomes smeared at higher $T$, 
it may be an enhancement of the upper edge due to the paramagnetic 
vortex core effect. 
We also see that the main peak shifts to higher $M$ in $P(M)$ 
with increasing $T$,    
since $M_{\rm para}({\bf r})$ outside of vortex [around position U] 
is shifted to higher. 

The experimental observation of the NMR resonance line shape 
was used to identify the FFLO phase 
in ${\rm CeCoIn_5}$.~\cite{kakuyanagi}  
There, the NMR spectrum shows the double peak structure 
in the FFLO phase, appearing new peak 
in addition to the main peak in the vortex state. 
This new peak is considered as a contribution of the FFLO nodal plane. 
This double peak structure is qualitatively reproduced by our calculation. 
In this status, our results may be  
qualitatively compared to the experimental data, 
since we did not perform fine tunings of parameters such as $L$, 
for the reason that we need further heavy calculations. 
The mechanism of rf-field penetration in the FFLO state~\cite{kakuyanagi} 
may further modify the NMR spectrum from the original $P(M)$. 
In the experimental data, 
the height of the new peak increases at higher $T$ 
in the NMR spectrum.~\cite{kakuyanagi}  
This feature is qualitatively consistent with the $T$-dependence in 
our results shown in Fig.  \ref{fig:KST},  
while it is an open question how $L$ increases as a function of $T$. 
It indicates that,  
among the $T$- and $L$-dependences discussed in 
Figs. \ref{fig:KSL} and \ref{fig:KST}, 
the contribution of the $T$-dependence seems to be dominant than 
the $L$-dependence  in the increasing $T$ process.

\section{Neutron scattering} 
\label{sec:SANS}

\begin{figure}[tb]
\includegraphics[width=8.0cm]{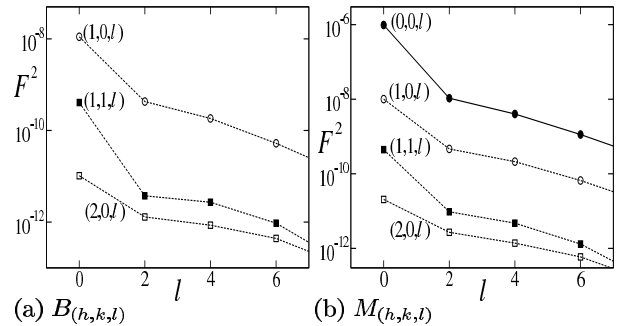} 
  \caption{
Factor $|F_{h,k,l}|^2$ for the intensity of the neutron scattering, 
calculated from the internal magnetic field $B_z({\bf r})$ (a) or 
from the paramagnetic moment $M_{\rm para}({\bf r})$ (b) 
in the $d$-wave pairing 
when the magnetic field is applied to the anti-nodal direction.   
$|F_{h,k,l}|^2$ is plotted as a function of $l$ for 
$(h,k)=(1,0)$, $(1,1)$ and $(2,0)$. 
In (b), we also show $|F_{h,k,l}|^2$ for $(h,k)=(0,0)$.
$\bar{B}=0.2 B_0$, $T=0.1T_c$ and $L=70R_0$. 
Since the period of the FFLO modulation is $L/2$ for $B_z({\bf r})$ 
and $M_{\rm para}({\bf r})$ along the $z$ direction, 
the diffraction peaks appear at spots with even $l$. 
 }
  \label{fig:SANS}
\end{figure}

The modulation of the internal magnetic field $B_z({\bf r})$ 
may be observed by the neutron scattering. 
If the periodic modulation along the $z$-direction is observed, 
it can be direct evidence of the FFLO modulation. 
Therefore, we discuss the neutron scattering 
in the FFLO vortex state. 
The intensity of the $(h,k,l)$-diffraction peak is given by 
$ I_{h,k,l}=|F_{h,k,l}|^2/|q_{h,k,l}| $ 
with the wave vector ${\bf q}_{h,k,l}$ given in Eq. (\ref{eq:q123}).
Here we write $(m_1,m_2,m_3)=(h,k,l)$ following notations  
of the neutron scattering. 
The Fourier component $F_{h,k,l}$ is given by 
$B_z({\bf r})=\sum_{h,k,l}F_{h,k,l}
\exp({\rm i}{\bf q}_{h,k,l}\cdot{\bf r})$. 
In Fig. \ref{fig:SANS}(a), we present 
$l$-dependence of $|F_{h,k,l}|^2$ for 
$(h,k)=(1,0)$, $(1,1)$ and $(2,0)$. 
The spot at $(h,k,l)=(1,0,0)$ and $(0,1,0)$ is used 
to determine the configuration and the orientation of the vortex lattice 
in the experiment of the small angle neutron scattering (SANS), 
and the higher component $F_{h,k,0}$ is used to 
estimate the structure of the internal magnetic field 
$B_z({\bf r})$.~\cite{kealey} 
It is noted that $F_{0,0,0}=\bar{B}$ and 
$F_{0,0,l}=0$ for $l\ne 0$, because average flux density 
$\bar{B}$ within the unit cell of the vortex lattice 
is constant along the $z$-direction. 
Therefore, to detect the FFLO modulation, 
we have to use the spot $(1,0,2)$, which has largest intensity 
among the spots related to the FFLO modulation. 
The spot $(1,0,2)$ is next to the spot $(1,0,0)$, 
which is used in the conventional SANS experiment 
to observe the stable vortex lattice configuration.

If the spins of the paramagnetic moment $M_{\rm para}({\bf r})$ 
strongly couple to the lattice modulation, 
it is possible to observe the structure of the paramagnetic moment 
through the lattice modulation by the neutron scattering. 
In this case, the Fourier component of the neutron scattering 
is given by 
$M_{\rm para}({\bf r})=\sum_{h,k,l}F_{h,k,l}
\exp({\rm i}{\bf q}_{h,k,l}\cdot{\bf r})$. 
In Fig. \ref{fig:SANS}(b), we present $|F_{h,k,l}|^2$ 
calculated from $M_{\rm para}({\bf r})$.  
Since $F_{0,0,l}$ ($l \ne 0$) is not zero in this case, 
the intensity at $(0,0,l)$ is larger than that at $(1,0,l)$ 
for the same $l$.

\section{Summary and discussion}

We investigated 
roles of the vortex and the FFLO nodal plane 
in the FFLO state, based on the quasiclassical Eilenberger theory. 
We selfconsistently calculated fully 3D spatial structure 
of the pair potential, 
the internal magnetic field, the paramagnetic moment, 
and local electronic states in the vortex lattice state 
under given period of the FFLO modulation $L$. 
We have seen that the topological structure of the pair potential 
determines the qualitative structure in the FFLO vortex state. 
At the FFLO nodal plane or at the vortex core, 
the $\pi$-phase shift of the pair potential 
gives rise to the sharp peaks in the local density of states, 
and enhances the local paramagnetic moment.  
Based on these results, we also discuss the experiments of 
NMR spectrum and neutron scattering, 
to identify characteristic 
behaviors in the FFLO states.  

In this work, we calculate the FFLO vortex structure 
both in the isotropic $s$-wave paring and in the anisotropic 
$d$-wave pairing with line nodes, 
when the FFLO nodal planes are perpendicular to the vortex direction.  
If we compare the FFLO vortex states 
under the same FFLO modulation period, 
we see that even in the $d$-wave pairing 
the qualitative structures of the FFLO vortex state are 
the same as in the $s$-wave pairing, and 
do not significantly depend on the relative angle between   
the applied magnetic field direction and the node direction 
within the $ab$ plane. 
This is because qualitative features reported in this paper 
come from the topological structure of $\Delta({\bf r})$. 
Quantitatively, the dependences of the FFLO vortex state 
on the pairing symmetry or the magnetic field direction relative to 
the node of the superconducting gap are important features 
to characterize the FFLO vortex state.\cite{ikeda,miranovic,adachiM}  
For the purpose, we need estimates of the optimized FFLO period $L$ 
and the phase diagram of the FFLO state in the mixed state. 
These calculations belong to future works  
within our framework of the selfconsistent calculations  
for the 3D structure of the FFLO vortex states.
However, qualitative features do not significantly depend on 
this process. 
We did some calculations for different $L$, 
and confirmed that our results reported in this paper  
are qualitatively unchanged as for the FFLO vortex structure, 
because they come from the topological structure of the FFLO states,  
as discussed above. 

Compared to results in the previous work by BdG theory,~\cite{mizushima2} 
we obtain qualitatively 
the same features of the paramagnetic moment and local electronic states 
around vortices and FFLO nodal planes 
in the quasiclassical calculation.  
We note that these qualitative features by 
the $\pi$-phase shift of the pair potential were confirmed  in this study 
also for the cases of $d$-wave pairing and the Q2D Fermi surface 
with parallel applied fields along $ab$ plane, 
which is not considered in the previous work 
by BdG theory assuming cylindrical symmetry structure and 
open boundary conditions. 
In this quasiclassical calculation with periodic boundary condition 
for the vortex lattice and FFLO modulation, 
we can take account of contributions by overlapping of 
electronic states from neighbor vortices or FFLO nodal planes. 
For example, we can discuss local structure 
at the mid-point between neighbor vortices. 
Compared with the previous quasiclassical work,\cite{tachiki} 
since we calculate fully 3D structure of the FFLO vortex states 
without mapping to 1D problem, we can explicitly analyze 
the contribution of the vortex lines in the FFLO vortex states. 
Thus, we can quantitatively estimate the spatial structure around vortices 
along the radial direction, 
including the vortex structure within the FFLO nodal plane. 
There, we see exotic structure at the intersection point of vortex 
and FFLO nodal plane. 
By this 3D calculation of the FFLO vortex states, 
we can evaluate the NMR spectrum and form factors of neutron scattering, 
including vortex contributions in the FFLO states.  

From the spatial structure of the paramagnetic moment 
in the FFLO vortex state, we can estimate the NMR spectrum 
in the Knight shift experiment.  
In the FFLO state, we obtain double peak structure 
of the NMR spectrum, as observed in the FFLO state of  
${\rm CeCoIn_5}$,~\cite{kakuyanagi} 
and discuss the dependences on the temperature and on the FFLO period.  
From the spatial structure of the internal magnetic field, 
we also discuss the neutron scattering in the vortex states 
as methods to detect the FFLO structure. 
Since this is a direct observation of the FFLO modulation, 
we hope that this feature will be observed 
in experiments to confirm the FFLO states.

\begin{acknowledgments}

The authors thank Y. Matsuda and K. Kumagai 
for valuable discussions and information on their experiments. 

\end{acknowledgments}


\end{document}